# High fidelity generation of complex optical field through scattering medium with iterative wavefront optimization

**Hui Liu,[1] Xiangyu Zhu, [1]Xiaoxue Zhang,[1] Xudong Chen,[1,2] Zhili Lin[1,3]**

[1]*Fujian Key Laboratory of Light Propagation and Transformation, College of Information Science and Engineering, Huaqiao University, Xiamen, Fujian 361021, China*
[2]*chenxd@hqu.edu.cn*
[3]*zllin2008@gmail.com*

**Abstracts:** Light scattering within scattering media presents a substantial obstacle to optical transmission. A speckle pattern with random amplitude and phase distribution is observed when coherent light travels through strong scattering media. Fortunately, wavefront shaping has been successfully employed with a spatial light modulator to recover intensity targets after scattering media, such as a sharp focus point or specified two-dimensional patterns. There have, however, been few studies that attempted to separately manipulate the amplitude and phase of the focusing field. In this study, a feedback-based wavefront shaping method to generate complex optical fields through scattering medium is proposed. A reliable phase retrieval approach is introduced to provide the complex feedback information, i.e., the amplitude and phase of the focusing field. Accordingly, in order to modulate the speckle field into a desired complex structured optical field, a multi-objective genetic algorithm is used to find the best phase map. To demonstrate the method's high performance, experimental tests have been carried out. High fidelity is demonstrated in the generation of diverse complex light fields, both in amplitude and phase. Our findings may facilitate the manipulation of light field through scattering medium, and are anticipated to further promote future applications such as optogenetics, vortex optical communication, and optical trapping through scattering media.

## 1.Introduction

Coherent light is transformed into a speckle field by multiple scattering and random interference between scattered-light components in disordered media, such as biological tissues and multimode fibers (MMF). This phenomenon poses a fundamental limitation to numerous research fields such as biomedical imaging, photodynamic therapy, and telecommunications. Fortunately, wavefront shaping technologies are now available to offer flexible solutions to minimize the influence of light scattering [1]. Different intensity targets, such as, a two-dimensional pattern or Gaussian focus spot, have been recovered after scattering medium by modulating the incident light into a special wavefront with a spatial light modulator (SLM). Transmission matrix (TM) inversion [2-4], feedback based iterative optimization [5-9], and optical time reversal/digital optical phase conjugation (DOPC) [10-14] are the most notable wavefront shaping approaches that have been proposed so far. With these techniques, modulated light with a special wavefront is transformed into a bright optical focus point or a desired focus pattern after the scattering medium.

The TM inversion method determines the optimal incident wavefront to focus the scattered light against the scattering medium by taking a measurement of the scattering medium's TM and solving an inverse problem [2]. The DOPC method is quick and non-iterative and it can focus in the millisecond range [11]. However, in order to accurately determine the phase of the scattered field and produce the conjugated wavefront, extremely precise system alignment and calibration are always required [15]. Unfortunately, the DOPC system's defects, including optical aberration, mechanical misalignment and even the curvature of the SLM itself, impose unavoidable limitations on the quality of DOPC in a complicated and coupled way [16]. Compared with the conventional TM methods that require an enumerate search, iterative optimizations can adjust the incident wavefront with respect to the current status of the dynamic scattering medium and therefore have the potential to be applied in highly dynamic environments [17, 18]. The anti-noise capability is, accordingly, a stronger strength of iterative optimization particularly genetic algorithm (GA) [7, 19, 20]. Beyond that, the global optimization ability of iterative optimization makes it be a general tool to benefit a broader community in wavefront shaping. By integrating feedback-based wavefront shaping with direct TM measurement, a recently developed feedback-assisted TM measurement technique has shown to reduce the required number of measurements [21]. Even more, particle swarm optimization (PSO) has recently been incorporated into DOPC to increase DOPC quality and facilitate the development of a high-performance DOPC system [14].

Leveraging the global optimization capability, traditional feedback-based iterative optimization effectively forces the scattering filed to form a single spot focusing or multi-point focusing. However, generation of complex optical field through scattering medium with iterative wavefront optimization has not been explored. Up to now, there have been some researches on the regeneration of vortex beams through scattering media utilizing TM or DOPC [22-26]. By TM inversion or digital filtering in the Fourier domain of the measured TM, conventional optical vortices [22-24] and perfect optical vortices [25] have been created. Based on DOPC, the generation of vortex beams through an MMF has also been demonstrated [26]. However, the quality of the regenerated structured beams in these reports are still an imperfection. In fact, unlike intensity-only target shaping, complex optical field generation through scattering media necessitates taking into account not only the intensity enhancement but also the precise distribution of the focus region's complex amplitude. Consequently, the generation of complex structured optical fields through scattering media is essentially a more challenging optimization process, and is susceptible to noise. It is anticipated that, in some ways, feedback-based iterative optimization has more elegant skills in providing higher quality.

In this study, we propose a complex feedback-based iterative optimization technique for the generation of structured optical fields through scattering medium. With a multi-objective optimization algorithm and complex feedback information obtained via phase retrieval, we experimentally demonstrate the merits of the proposed method with generation of a range of complex optical fields. High-fidelity generation of complex optical fields in both amplitude and phase are presented.

## 1. Principle and method

*2.1 The model of complex optical field shaping through scattering medium*

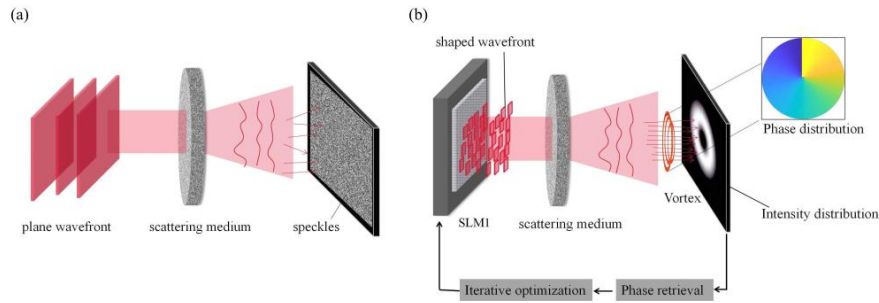

Fig. 1. Concept diagram of generation of complex optical field through scattering medium with iterative wavefront optimization. (a) A plane wave is focused on a scattering medium, and a speckle pattern is transmitted. (b) The wavefront of the incident light is shaped to generate a desired complex filed at target area.

Figure 1 shows the principle of the experiment. Normally, incident light is scattered by the scattering medium and forms a random speckle pattern [Fig. 1(a)]. The goal is to match the incident wavefront to the sample so that a complex optical field with desired amplitude and phase distribution is generated in a specified target area. For this purpose, a closed-loop feedback system incorporated with a phase retrieval method is proposed. As shown in Fig. 1(b), input light can be modulated with a SLM (SLM1). A phase retrieval method collects the output light after scattering medium, and estimates the complex amplitude of the speckle field. According to the complex feedback including amplitude and phase information, iterative optimization algorithm is used to search for the optimal phase mask to optimize the input light. Finally, a desired structured beam such as vortex is obtained at the target area.

*2.2 The principle of phase retrieval*

In this study, to estimate the complex amplitude distribution of speckle field, a phase retrieval method called WISH is adapted from [27], which measures wavefront computationally using sequential captures with a second SLM (SLM2). As shown in Fig. 2, a wavefront imaging sensor (WIS) consists of SLM2, a camera and a phase retrieval algorithm. The scattered light after scattering medium, with unknown wavefront of $u$, incidents on the SLM2 plane in WIS. The incident wavefront is then encoded with pre-designed random phase patterns sequentially by SLM2, while the corresponding intensity are collected by camera. Next, the unknown field $u$ is computed with a Gerchberg-Saxton (GS) algorithm [28] by alternating projections between SLM2 and the camera plane.

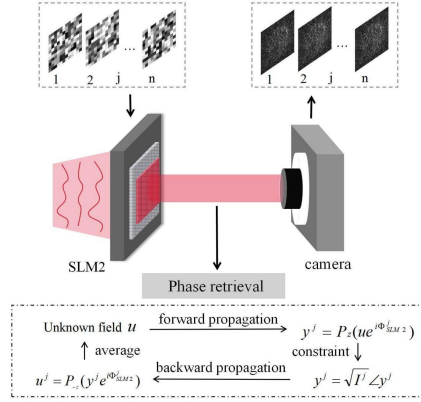

Fig. 2. The concept diagram of WIS. $\Phi_{SLM2}^j$ is the $j_{th}$ phase modulation on SLM2 and $I^j$ is the corresponding recorded intensity.

*2.3. Multi-objective optimization algorithm*

As previously mentioned, in complex optical field focusing optimization, two objectives need to be addressed: the amplitude and phase distribution of the focusing field. To assess the fidelity of the generated complex field, we propose utilizing two discriminants, $\gamma$ and $\delta_p$. $\gamma$ is the Pearson Correlation Coefficient (PCC) between the target and the evaluated intensity obtained from phase retrieval, calculated over the region of interest (ROI) [29]. $\delta_p$ is the relative error between the target phase and the measured one, evaluated over non-zero amplitude within the ROI. Generally, a larger $\gamma$ and a smaller $\delta_p$ corresponds to a complex focusing optical field with higher fidelity.

In fact, iterative optimization of complex optical field generation requires simultaneous shaping of the speckle field's amplitude and phase. As a result, it is a multi-objective optimization (MOO) problem [30]. Thus it cannot be perfectly solved with a single-objective optimization algorithm like GA [19] or PSO [14], in a sense. As one of the best MOO algorithms, Non-dominated Sorting Genetic Algorithm II (NSGA-II) has been successfully applied to achieve high enhancement and acceptable uniformity in multi-point light focusing [31], and to generate controllable spectrum at various positions from speckle patterns [32]. In this study, we employ NSGA-II to achieve inverse diffusion for complex optical field focusing. We note that, $\gamma$ and $1-\delta_p$ are used as the two fitness functions. They can operate effectively for a variety of complex optical fields, including those with non-binary amplitude and phase distributions.

## 2. Experimental setup and results

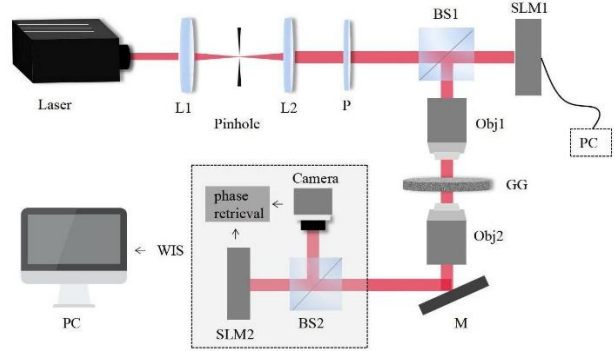

Fig. 3. The experimental setup of complex optical field generation through scattering medium. L1-L2, lenses; P, polarizer; BS1-BS2, non-polarizing beam splitter; Obj1-Obj2, objective lenses; GG, ground glass; M: mirror; SLM1:SLM2, spatial light modulators; PC, personal computer.

We verify the proposed method experimentally with the setup shown in Fig. 3. The laser beam (1064 nm) is filtered and expanded by a telescope formed by lens L1, pinhole and lens L2. A polarizer (P) is used to produce a horizontally polarized laser beam. Then, the incident light is shaped and reflected by SLM1 (Holoeye PLUTO NIR). In SLM1, 35×35 input modes are used with 6×6 pixels grouped as a mode. The modulated light is reflected by the beam splitter (BS1) and focused by the objective (Obj1, 40 ×, NA=0.65) into the ground glass (GG, Thorlabs DG10-220, 220 grit, 2 mm thickness). Another objective (Obj2, 25 ×, NA=0.4) is placed behind the GG to collect the scattered light. The output scattered light is then reflected by a mirror (M) and transmitted to the WIS. In WIS, another BS (BS2) guides the incoming scattered light into SLM2 (Hamamatsu X13138-03WR). SLM2 encodes the unknown diffused light with multiple predefined patterns sequentially, while the camera (Thorlabs, CS2100M) at a distance of 284 mm from SLM2 records the corresponding intensity. Based on the predefined patterns and the corresponding intensity data, a GS algorithm estimates the complex information of the field at target plane. Finally, according to the complex information obtained from WIS, the iterative optimization algorithm searches for the optimal phase mask to generate the desired complex field at target plane. During the experiment, control of the SLM1, SLM2 and camera is performed by the PC.

We created a range of target optical fields to verify the proposed approach, as shown in Fig. 4. The corresponding experimental results are shown in Fig. 5. It can be seen that complex optical fields with various amplitude and phase distribution have been generated with high quality. The strong correlation between the experimental and the corresponding targets indicates the excellent capability of the proposed method. These impressive results are achieved for complex fields with intensity distribution, whether discrete or continuous. Besides, the corresponding fitness function values of the final results are listed in Table 1. The high values of $\gamma$ and $1-\delta_p$ further confirm that we have generated these complex structured optical field with high fidelity.

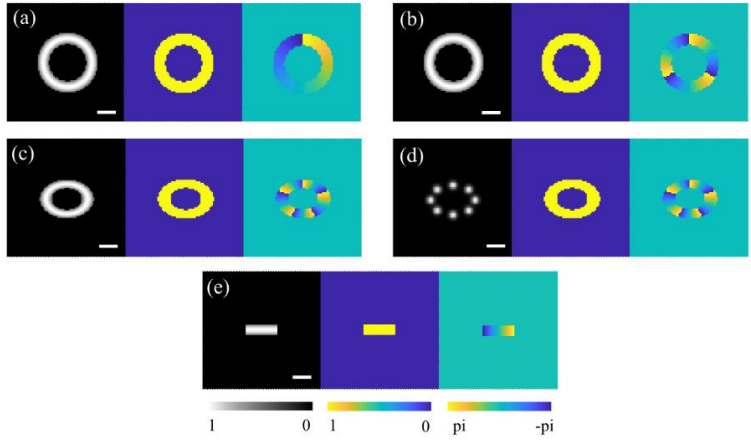

Fig. 4. Target complex optical field to be generated through scattering medium. The left columns are normalized target intensity distributions; The middle columns are binary masked regions for target phase; The right columns are phase distributions in the region of binary masks. Optical vortices (OVs) with topological charge (TC) (a) $l=1$; (b) $l=3$; (c) Elliptical optical vortex (EOV) with TC $l=5$; (d) Lattice-EOV; (e) Gaussian line. The scale bar represents 50μm.

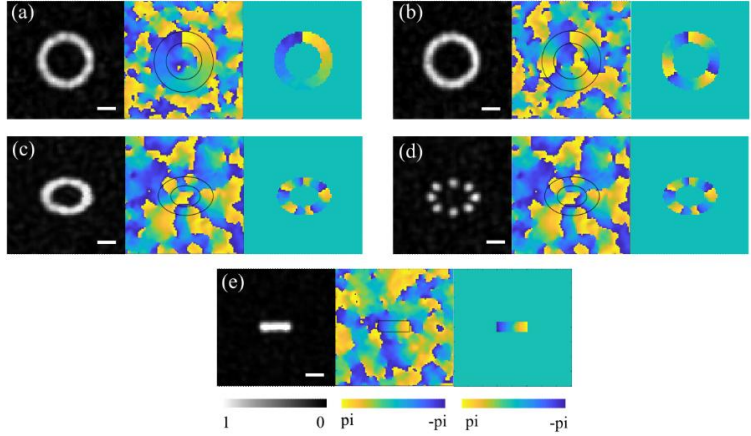

Fig. 5. The experimental results of complex optical field generation through scattering medium. The left columns are normalized estimated intensity distributions in the ROI; The middle columns are phase distributions in the ROI, the intersection region of black solid lines in each middle column indicates the binary mask; The right columns are phase distributions in the region of binary mask. The scale bar represents 50μm.

**Table 1. The optimization results of the patterns in Fig 5.**

| patterns | fidelity | | NMSE | $N$ | $\eta_{AVG}$ | std |
|---|---|---|---|---|---|---|
| | $\gamma$ | $1-\delta_p$ | | | | |
| (a) OV$(l=1)$ | 0.9851 | 0.9796 | 0.0621 | 772 | 8.5300 | 0.1201 |
| (b) OV$(l=3)$ | 0.9753 | 0.9873 | 0.0585 | 772 | 7.9778 | 0.1177 |
| (c) EOV$(l=5)$ | 0.9524 | 0.9487 | 0.0806 | 504 | 10.2181 | 0.1338 |
| (d) Lattice-EOV | 0.8873 | 0.9313 | 0.0932 | 504 | 5.1655 | 0.2250 |
| (e) Gaussian line | 0.9768 | 0.9896 | 0.0556 | 147 | 10.8467 | 0.1265 |

## 3. Discussion

*4.1 Discussion on the performance of WIS and quality of the generated complex fields*

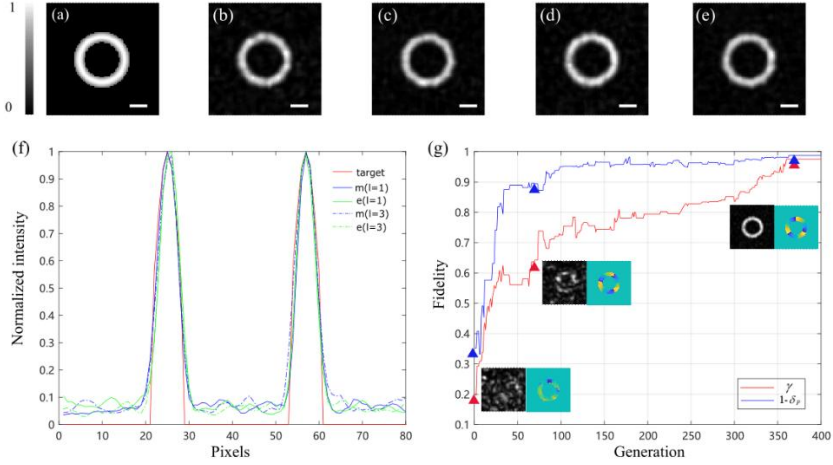

Fig. 6. (a) Target intensity of OVs; (b) Measured and (c) Estimated intensity of OV $(l=1)$; (d) Measured and (e) Estimated intensity of OV $(l=3)$; (f) Center horizontal intensity profiles of Fig. 6. (a)-(e); (g) The optimization process of OV $(l=3)$ in Fig. 5(b). The scale bar represents 50μm.

We note that the complex feedback information from phase retrieval is critical to the entire iterative wavefront optimization process. Once the optimization progress had converged, the optimal phase mask was displayed on SLM1, and a blank pattern with uniform grayscale valued of 0 was imposed on SLM2, in order to obtain the corresponding ground truth of final results in the ROI, which is regarded as measured intensity and recorded by the camera. Fig. 6(a)-(e) presents the target intensity patterns for OVs $(l=1,3)$, the corresponding measured and estimated intensity of the final optimization results, respectively. Compared with Fig. 6(b)-(c) and Fig. 6(d)-(e), it is obvious that the difference between the normalized measured intensity and the estimated one is slight. In addition, the horizontal intensity profiles of Fig. 6(a)-(e) are also plotted in Fig. 6(f). It is evident that the central horizontal intensity profiles of the target intensity, the corresponding measured and estimated intensity of the final optimization results almost overlaps with each other. Furthermore, the accuracy of phase retrieval can be quantitively evaluated by the normalized mean squared error (NMSE) [33, 34]. As shown in Table 1, the low values of NMSE for each pattern exhibits a relatively low error. In general, the high accuracy of phase retrieval is confirmed.

Besides, high intensity uniformity can also be reflected from the high values of PCC. In fact, the intensity uniformity of the focal points is also one of the crucial aspects for the generation of complex field through scattering media. It should be emphasized that, as shown in Fig. 5, the uniformity of all the intensity distribution in our experimental results is also remarkable. Compared with the results in [23, 25, 26], the OVs $(l=1,3)$ generated in our work exhibit better intensity uniformity. To further assess the intensity uniformity quantitively, the standard deviation (std) of intensity is evaluated [31]. As shown in Table 1, despite the huge number of focal points, the evaluation results of std demonstrate the outstanding performance of the proposed method on uniform intensity focusing.

As one of the most important indicators of a wavefront shaping system's focusing ability through scattering media, the average enhancement factor $\eta_{AVG}$ for the experimental results in

Fig. 5 is also further examined [29]. As listed in Table 1, even though the total enhancement is considerable, the average enhancement is rather modest, due to the huge number $N$ of focal points in the focal region. It can be further increased by increasing the number of control elements of SLM1 [35].

*4.2 Discussion on the efficiency*

In this section, we will discuss the efficiency of the method we proposed. In our experiments, the population size is 60, and the offspring size is 30. Since, 8 patterns are required to accurately estimate the complex information by WIS, typically 240 measurements are required per generation. We note that after the end of 240 measurements, all the recorded data for phase retrieval is calculated in parallel on GPU (NVIDIA GeForce RTX 3900) for acceleration. The process of one generation in our experiments takes about 9.8 s, typically. As an instance the optimization progress of OV $(l=3)$ is shown in Fig. 6(g), it converged after about 360 iterative cycles in 58.8 minutes. In this work, not only 772 focus points with high uniformity are obtained, but also a precise phase distribution of the focus field is optimized simultaneously. For comparison, in the previous reports for multi-point focusing through scattering medium based iterative wavefront optimization [31], the experiment for 20 points focusing took about 48.5 minutes.

## 4. Conclusion

In summary, we have first presented a powerful in situ method for generation of complex optical fields after strong scattering medium. This approach extends the traditional feedback-based wavefront shaping from intensity-only focusing to complex optical field generation. It has shown high-fidelity performance in a range of optical fields after strong scattering medium in the experiment. With the aid of the complex feedback and a multi-objective genetic optimization algorithm, our approach makes it possible to accomplish promising application such as micromanipulation and OAM communication in complex environments that introduce strong aberrations in the laser beams.

**Funding.** National Natural Science Foundation of China (61605049); The Youth Innovation Foundation of Xiamen City (3502Z20206013); The Fundamental Research Funds for the Central Universities (ZQN707).

**Disclosures.** The authors declare no conflicts of interest.

**Data availability.** Data underlying the results presented in this paper are not publicly available at this time but may be obtained from the authors upon reasonable request.